\documentclass[unnumsec,webpdf,contemporary,large]{oup-authoring-template}%

\usepackage{textcomp}
\graphicspath{{Fig/}}
\usepackage{subfig}
\usepackage{dsfont} 

\theoremstyle{thmstyleone}%
\theoremstyle{thmstyletwo}%
\theoremstyle{thmstylethree}%

\begin{document}

\journaltitle{Briefings in Bioinformatics}
\DOI{DOI HERE}
\copyrightyear{2022}
\pubyear{2022}
\access{Advance Access Publication Date: Day Month Year}
\appnotes{Paper}

\firstpage{1}


\title[Attention-wise masked graph contrastive learning]{Attention-wise masked graph contrastive learning for predicting molecular property}

\author[1]{Hui Liu}
\author[2]{Yibiao Huang}
\author[1]{Xuejun Liu}
\author[2,$\ast$]{Lei Deng}

\authormark{Huang et al.}
\address[1]{\orgdiv{School of Computer Science and Technology}, \orgname{Nanjing Tech University}, \orgaddress{ \postcode{211816}, \state{Nanjing}, \country{China}}}
\address[2]{\orgdiv{School of Computer Science and Engineering}, \orgname{Central South University},\orgaddress{\postcode{410075}, \state{Changsha}, \country{China}}}

\corresp[$\ast$]{Correspondence should be mainly addressed to \href{email:leideng@csu.edu.cn}{leideng@csu.edu.cn}.}

\received{Date}{0}{Year}
\revised{Date}{0}{Year}
\accepted{Date}{0}{Year}



\abstract{\textbf{Motivation:} Accurate and efficient prediction of the molecular properties of drugs is one of the fundamental problems in drug research and development. Recent advancements in representation learning have been shown to greatly improve the performance of molecular property prediction. However, due to limited labeled data, supervised learning-based molecular representation algorithms can only search limited chemical space, which results in poor generalizability.  \\
\textbf{Results:}  In this work, we proposed a self-supervised representation learning framework for large-scale unlabeled molecules. We developed a novel molecular graph augmentation strategy, referred to as attention-wise graph mask, to generate challenging positive sample for contrastive learning. We adopted the graph attention network (GAT) as the molecular graph encoder, and leveraged the learned attention scores as masking guidance to generate molecular augmentation graphs. By minimization of the contrastive loss between original graph and masked graph, our model can capture important molecular structure and higher-order semantic information. Extensive experiments showed that our attention-wise graph mask contrastive learning  exhibit state-of-the-art performance in a couple of downstream molecular property prediction tasks.
}

 \boxedtext{
 \begin{itemize}
 \item An novel attention-wise method for graph augmentation was developed to generate challenging positive sample for contrastive learning. In applying to molecular property prediction, our method achieved better performance than fully-supervised model and multiple benchmark methods.
 \item The contrastive view augmented by masking a percentage of nodes (edges) with high attention weights yielded better performance, indicating that attention-guided contrastive views were greatly helpful to learn expressive molecular representation.
 \item Pretraining on larger molecule set showed better performance on various downstream tasks, thereby we drew conclusion that our model had capacity for large-scale dataset to obtain better generalization.
 \item Visualization of attention weights implied that the nodes essential to downstream tasks were accordingly given higher attention weights.
 \end{itemize}}

\maketitle

\section{Introduction}
The physicochemical properties of molecules, such as water solubility, lipophilicity, membrane permeability and degree of dissociation, are of great importance to the screening of leading compounds. Traditional wet-lab experiments is time-consuming and labor-intensive, which is impossible to cover hundred millions of candidate molecules \citep{RN1}. So, many computational methods have been proposed to predict molecular properties. There methods drastically promoted the efficiency of drug discovery and return on investment, so that accelerated the drug discovery process \citep{RN2}.

Molecular representation is crucial to identify various properties of molecules \citep{RN3,RN4,RN5}. The feature engineering-based chemical fingerprint, in which each bit represents the absence or presence of a certain biochemical property or substructure, transforms the structural information or physicochemical properties of the molecule to a fixed-length vector \citep{RN6}. For example, PubChem fingerprint \citep{RN7} and extended connectivity fingerprints (ECFPs) \citep{RN8} are frequently used representations of molecules. However, most chemical fingerprints rely on expert knowledge, and include only task-specific information, and lead to limited performance when applied to other tasks. In recent years, deep learning has achieved remarkable success in natural language processing (NLP) \citep{RN9}, computer vision (CV) \citep{RN10}, and graph structure prediction \citep{RN11}. Many studies have applied deep learning to chemical modeling \citep{RN12,RN13,RN14,RN15,RN16,RN17}, and drug discovery \citep{RN18,RN19,RN10}, etc. However, the performance of deep learning depends on a large amount of manually labeled samples  \citep{RN13,RN21}, such as molecules with known properties in our case. When applied to small-size datasets, the deep learning model is vulnerable to overfitting, which leads to poor generalizability.

In recent years, self-supervised learning has caught much attention because of its better generalizability in multiple fields. Self-supervised learning first run pretraining process on large-scale unlabeled dataset to derive latent representations, and then applied to downstream tasks through transfer learning \citep{RN9,RN10,RN22} to obtain better performance and robustness  \citep{RN23,RN24}. For molecular property prediction task, a few self-supervised methods have been proposed for molecular representation learning \citep{RN25,RN26,RN27,RN28,RN29,RN30}. These methods fall roughly into two categories: generation-based methods and contrastive learning-based methods. The generative methods learned molecular features by establishing specific pretext tasks that encourage the encoder to extract high-order structural information. For example, MG-BERT learned to predict the masked atomic \citep{RN25}, by integrating the local message passing mechanism of graph neural networks (GNNs) into the powerful BERT model \citep{RN26} to enhance representation learning from molecular graphs. MolGPT \citep{RN27} trained a transformer-decoder model for the next token prediction task using masked self-attention to generate novel molecules. Contrastive representation learning encourages augmentations (views) of the same molecules to have more similar representations compared to augmentations generated from different molecules. For example, MolCLR \citep{RN28} proposed three different graph augmentation by masking nodes or edges or subgraphs, and then maximize the agreement of the augmentations from same molecule while minimizing the agreement of different molecules. CSGNN \citep{RN29} designs a deep mix-hop graph neural network to capture higher-order dependencies and introduces a self-supervised contrastive learning framework. MolGNet \citep{RN30} uses both paired subgraph recognition (PSD) and attribute masking (AttrMasking) to achieve node-level and graph-level pre-training, which improves the ability to extract feature from molecular graphs.

The performance of contrastive representation learning often relies on the quality of augmented views. Current methods generated augmentation of molecular graphs by randomly masking some nodes and edges [28]. However, random mask cannot guide the encoder to capture the most important substructure. In this work, we propose an attention-wise contrastive learning framework for molecular representation and property prediction. Specifically, we first constructed the molecular graph from SMILES, and used the graph attention network (GAT) as encoder to transform molecular graph into latent representation. Next, we leveraged the attention weights of nodes and edges learned by GAT to generate the augmentation graph, by masking a percentage of nodes or edges according to their attention weights. By minimization of the contrastive loss between original graph and the challenging augmented graph, our model was driven to capture important substructure and higher-order semantic information, and thus produce concise and informative molecular representation. We conducted extensive experiments and showed that the molecular representations learned by our method exhibit state-of-the-art performance in multiple downstream molecular property prediction tasks. Performance comparison verified our method outperformed competitive methods. Moreover, we explored the interpretability and found the attention wights revealed importance patterns of important substructure. Finally, we applied our learning framework to ten millions of molecules and yielded their expressive representation ready for various downstream application.

\section{Materials and methods}
\subsection{Data source}
For pretraining, we downloaded two large-scale molecule datasets from ZINC substance channel. One is the in vitro set, which includes 306,347 SMILES descriptors of the substances reported or inferred bioactive at 10 $\mu M$ or better in direct binding assays. All molecules in the in vitro set were used in our model ablation experiments. The other is built from the ZINC now set, which means it include all in-stock and agent substances for immediate delivery. This dataset includes 9,814,569 unique SMILES descriptors. The open-source tool RDkit \citep{RN31} was used to transform each SMILES descriptor to molecular graph, in which each node represents an atom and edge represent a chemical bond. The molecular graphs were ready for the input of graph attention network.

For downstream performance evaluation, we chose 7 datasets from MoleculeNet \citep{RN32}, which collected more than forty molecular property prediction tasks. Table \ref{tab:1} showed the total number of molecules in each dataset, as well as the number of subtasks. They covered different molecular properties, including membrane permeability, toxicity, bioactivity. For each dataset, we use the scaffold split from DeepChem \citep{RN33} to create an 80/10/10 train/valid/test subset. Rather than random split, scaffold-split divided the molecules based on their substructures, making the prediction task more challenging yet realistic.
\begin{table}[]
\centering
\caption{Seven datasets representing prediction tasks of different molecular properties}
\label{tab:1}
\begin{tabular}{cccccccll} \hline
\multicolumn{1}{l}{DataSet} & \multicolumn{1}{l}{\ Molecule number} & \multicolumn{1}{l}{\ Task number} & \multicolumn{1}{l}{} & \multicolumn{1}{l}{} & \multicolumn{1}{l}{} & \multicolumn{1}{l}{} &  &  \\ \hline
BBBP & 2,039 & 1 & \textbf{} &  &  &  &  &  \\
BACE & 1,513 & 1 & \textbf{} &  &  &  &  &  \\
HIV & 41,127 & 1 & \textbf{} &  &  &  &  &  \\
ClinTox & 1,478 & 2 & \textbf{} &  &  &  &  &  \\
Tox21 & 7,831 & 12 &  &  &  &  &  &  \\
SIDER & 1,478 & 27 & \textbf{} &  &  &  &  &  \\
MUV & 93,087 & 17 &  &  &  &  &  & \\ \hline
\end{tabular}

\end{table}
\subsection{Method}
\subsubsection{ATMOL framework}
\begin{figure*}[htb]
	\centering
	\includegraphics[scale=0.65]{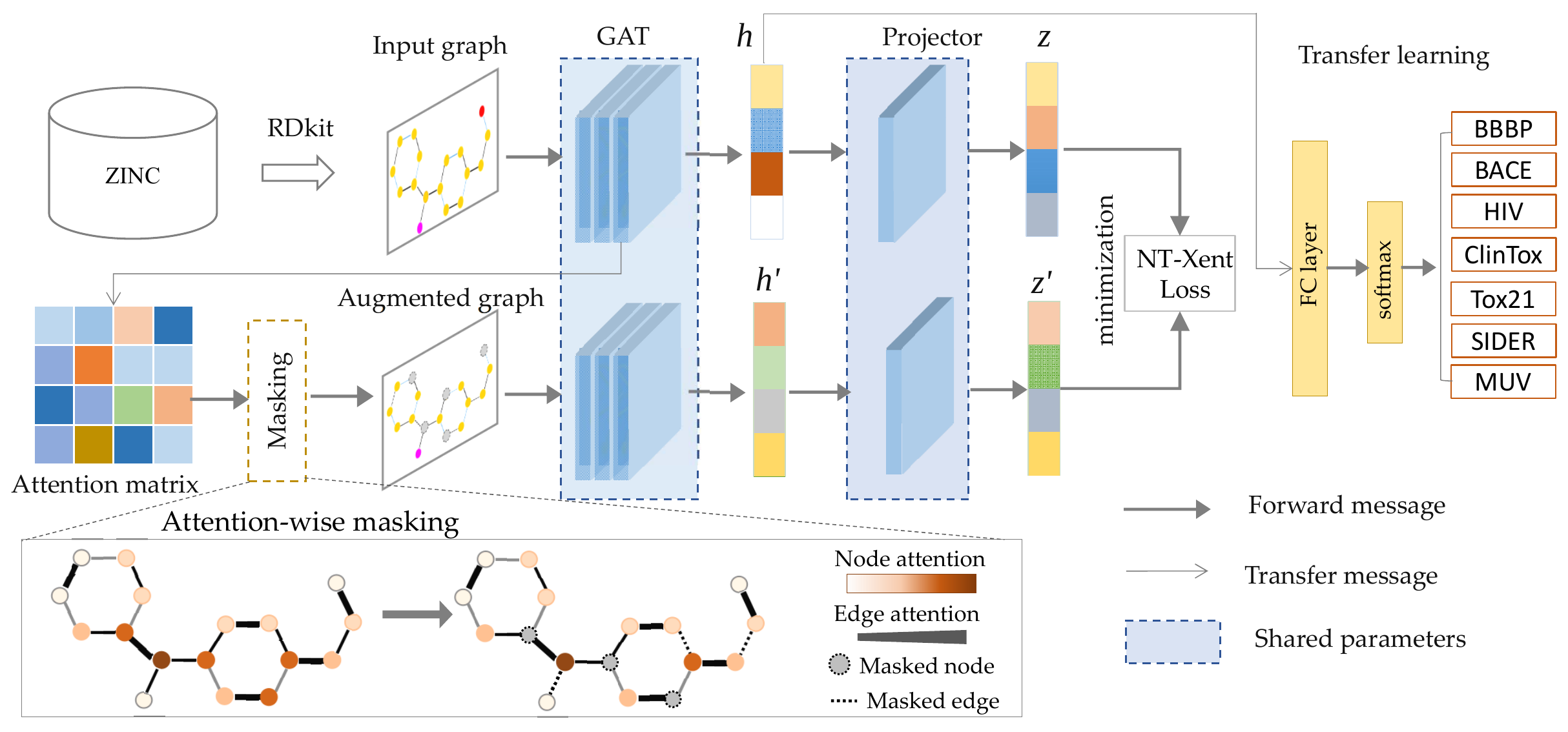}
	\caption{Illustrative flowchart of the proposed ATMOL contrastive learning framework for molecular property prediction. }
	\label{fig:flowchart}
\end{figure*}

Our framework consisted of two steps: pre-training and transfer learning. As shown in Figure \ref{fig:flowchart}, we first performed contrastive learning on large-scale unlabeled datasets to obtain molecular representations, and then apply transfer learning to predict molecule properties. The molecular graph was taken as input, and mapped to latent representation by GAT encoder. Meanwhile, an attention-wise masking module closely tracked the GAT encoder and use the attention scores to produce augmented view (masked graph) by masking some nodes or edges. We deliberately designed the masking module to produce augmented graph that posed a challenge for the encoder to generate similar representations of the graphs from same molecule but dissimilar representations from other molecules. Consequently, the contrastive learning model was forced to capture important substructure and higher-order semantic information, and thus produce concise and informative molecular representation. We validated the molecule representations obtained by contrastive learning on seven molecular property prediction tasks.

\subsubsection{Molecule graph embedding }
The graph attention network (GAT) is a multihead attention-based architecture developed to learn molecule graph embedding. The GAT architecture is built from multiple graph attention layers, and each layer applies linear transformation to node-level representations for calculating attention scores. Let $h_i$ be the embedding of node $i$, $W$ is a learnable attention weight matrix. The attention score $\alpha_{i,j}$ between node $i$ and its first-order neighbor node $j$ is calculated as
\begin{equation}\label{equ:1}
  \alpha_{ij}=\frac{\exp (elu(a^T(Wh_i,Wh_j)))}{\sum_{k\in N(i)}\exp (elu(a^T(Wh_i,Wh_k)))}
\end{equation}
where $a$ is a learnable vector, $elu$ is the exponential linear unit activation function. The attention score $\alpha_{ij}$ was actually the softmax normalized message between node $i$ and its neighbors. Once the attention scores were computed, the output feature of node $i$ is aggregating the neighbor features weighted by the attention scores:
\begin{equation}\label{equ:1}
  h_i=\sigma(a_{ii}Wh_i+\sum_{j\in N(i)}\alpha_{ij}Wh_j)
\end{equation}
where $\sigma(.)$ is the ReLU activation function. In our model, we used two GAT layers. The multi-head attention mechanism was applied to the first layer and the number of heads was set to 10. Given the node-level features, the global max pooling layer is used to obtain the graph embedding. The dimension of hidden features is set to 128.

\subsubsection{Attention-wise mask for graph augmentation}
To produce high-quality augmented graphs, we masked a percentage of nodes (edges) of the input molecular graph according to the attention scores learned by GAT encoder. Once a node (edge) was masked, its embedding was set to 0. If an edge was masked, the message passing along this edge was removed during the graph convolutional operation. Formally, we define the masking rate $r$, representing the percent of nodes (edges) of the input graph would be masked. We iteratively sampled nodes (edges) until the percentage of masked nodes (or edges) reached the predefined masking rate $r$. To explore the effect of masking different edges and nodes, we tried different masking strategies:
\begin{enumerate}[\hspace{0.5em}(a)]
  \item[1)] Max-attention masking: $r$\% nodes (edges) with the largest attention scores were masked. This masking produced an augmented view with greatest difference to the current view of the input molecule graph.
  \item[2)] Min-attention masking: $r$\% nodes (edges) with the smallest attention scores were masked. This masking produced an augmented graph with least difference to the current view of the input molecule graph.
  \item[3)] Random masking: $r$\% nodes (edges) was randomly selected from the input graph and masked. This masking neglected the attention scores and randomly masked a percentage of nodes (edges). This masking strategy was commonly used in previous studies, we included it for comparison.
  \item[4)] Roulette-masking: the probability of each node or edge to be masked was proportional to its attention score. The attention weight matrix $W$ was normalized by softmax function, and yielded a probability distribution. The probability of an nodes (or edge) being masked is proportional to the probability.
\end{enumerate}

In this study, the multihead attention weight matrices in the first graph layer were aggregated into an integrative attention matrix in the second layer, which was used in the attention-wise mask module to produced augmentation graph.

\subsubsection{Contrastive learning}
The GAT transformed the input molecule graph and its augmented graph into latent vectors $h_i$ and $h'_i$, which were then mapped to $z_i$ and $z'_i$ by a nonlinear projector. Next, the similarity $sim(z_i, z'_i)$ between two projected views was computed. We adopted the we normalized temperature-scaled cross entropy (NT-Xent) as the contrastive loss function:
\begin{equation}\label{equ:1}
 \mathcal{L}=\log \frac{\exp(sim(z_i,z'_i)/\tau)}{\sum_{k=1}^{2N}{1}_{[k\neq i]}\exp(sim(z_i,z'_i)/\tau)}
\end{equation}
where ${1}_{[k\neq i]}\in\{0, 1\}$ is an indicator function evaluating to 1 iff $ k\neq i$, $\tau$ denotes a temperature parameter; $N$ is the number samples in a mini-batch. In our study, the cosine distance was used to evaluate the similarity of two views of a same molecule.

Negative samples play a crucial role in self-supervised representation learning, and previous studies have confirmed that a large number of negative samples helped improving performance. So, apart from taking other samples of minibatch as negative samples, we added the augmented molecule graphs generated by attention-wise masking to the negative sample pool, so that the number of negative samples was greatly extended. Of more importance, the augmented molecule graph enriched the diversity of negative samples.

The advantages our attention-wise masking for graph augmentation were reflected in two aspects. First, attention-wise masking generated challenging positive sample pairs, which increased the difficulty of contrastive learning, thereby preventing to learn collapsed latent representation. Also, it enriched the diversity of negative samples that are helpful to escape from local optimal solution during the minimization of the contrastive loss.

\subsubsection{Pretraining and transfer learning}
In the pretraining stage, we conducted contrastive learning on large-scale unlabeled data to obtain molecular representation for each molecule. The GAT encoder used  10-head self-attention mechanism in the first graph layer and aggregated them by mean-pooling in the second layer. The contrastive loss is optimized using the batch gradient descent algorithm by Adam optimizer. The learning rate is set to 1e-4, the batch-size is set to 128, and the number of pre-training epochs is set to 20 epochs.
During the transfer learning for molecular property prediction, we appended two fully-connected layers to the GAT feature extraction network. We froze the weights of the GAT encoder and tuned only the two fully-connected layer in the fine-tuning stage. The cross-entropy loss was applied for all classification task, and the learning rate is set to 1e-7. Adam optimizer was used and batch-size was set to 100. The performance of molecular property prediction was evaluated by ROC-AUC. The early stopping and dropout strategies were used to prevent overfitting.

Each dataset used for molecular property prediction was split into training, validation and test datasets in a ratio of 8:1:1. The pretrained model was fine-tuned on the training set and validated on the validation set. To avoid random bias, the pretraining and validation process was repeated for five times and each time evaluated on the test set. The mean AUC values were reported as the final performance.

\section{Results}
\subsection{Contrastive learning boosted performance}
We first verified whether contrastive learning-based pre-training improved the performance on downstream tasks. For this purpose, we compare our method to the model without pre-training on different molecular property prediction tasks. For each task, we trained a fully-supervised model, and its architecture included a GAT encoder for molecular graph embedding and two fully-connected layers for molecular property prediction. To systematically evaluating the effectiveness of pre-training, we took into account different molecular augmented graphs produced by masking nodes, edges or both nodes and edges (masking ratio r was 25\%). Table \ref{tab:2} showed the ROC-AUC values of these competitive models on seven benchmark tasks. It can be seen that pre-training significantly boosted the performance on various molecular property prediction tasks, including membrane permeability, toxicity and bioactivity.
\begin{table*}[]
\centering
\caption{ROC-AUC (\%) values of pretrained models and fully-supervised model on seven molecular properties prediction tasks}
\label{tab:2}
\begin{tabular}{llllllll} \hline
Model & BBBP & BACE & HIV & ClinTox & Tox21 & SIDER & MUV \\ \hline
Fully-supervised & 85.5$\pm$1.4 & 76.2±0.1 & 72.6$\pm$0.1 & 92.6$\pm$0.2 & 76.2$\pm$0.1 & 80.5$\pm$0.2 & 69.8$\pm$0.1 \\ \hline
\begin{tabular}[c]{@{}l@{}}ATMOL (mask nodes)\end{tabular} & 91.1$\pm$0.4 & 84.3$\pm$0.6 & 75.1$\pm$0.1 & 97.1$\pm$0.1 & \textbf{79.1$\pm$0.2} & 81.4$\pm$0.3 & \textbf{79.1$\pm$0.1} \\
\begin{tabular}[c]{@{}l@{}}ATMOL (mask edges)\end{tabular} & 90.3$\pm$0.8 & 82.5$\pm$0.4 & 80.7$\pm$0.3 & 96.5$\pm$0.3 & 76.2+0.1 & 80.2+0.1 & 79.0+0.4 \\
\begin{tabular}[c]{@{}l@{}}ATMOL (mask nodes and edges)\end{tabular} & \textbf{92.1$\pm$0.1} & \textbf{87.3$\pm$0.3} & \textbf{81.2$\pm$0.5} & \textbf{97.5$\pm$0.3} & 77.1$\pm$0.3 & \textbf{81.9$\pm$0.2} & 78.8$\pm$0.1 \\ \hline
\end{tabular}

\end{table*}
Meanwhile, we found that the molecular graph augmented by masking both nodes and edges achieved better performance, compared to those augmented graphs by masking nodes or edges alone. The results showed that contrastive learning-based pre-training obtained informative molecular representations, and thereby greatly improved the performance on downstream tasks.

\subsection{Masking strategy affected feature extraction }
\begin{figure}[htbp]  
	\captionsetup{labelformat=simple, position=top}
	\centering
	\includegraphics[scale=0.75]{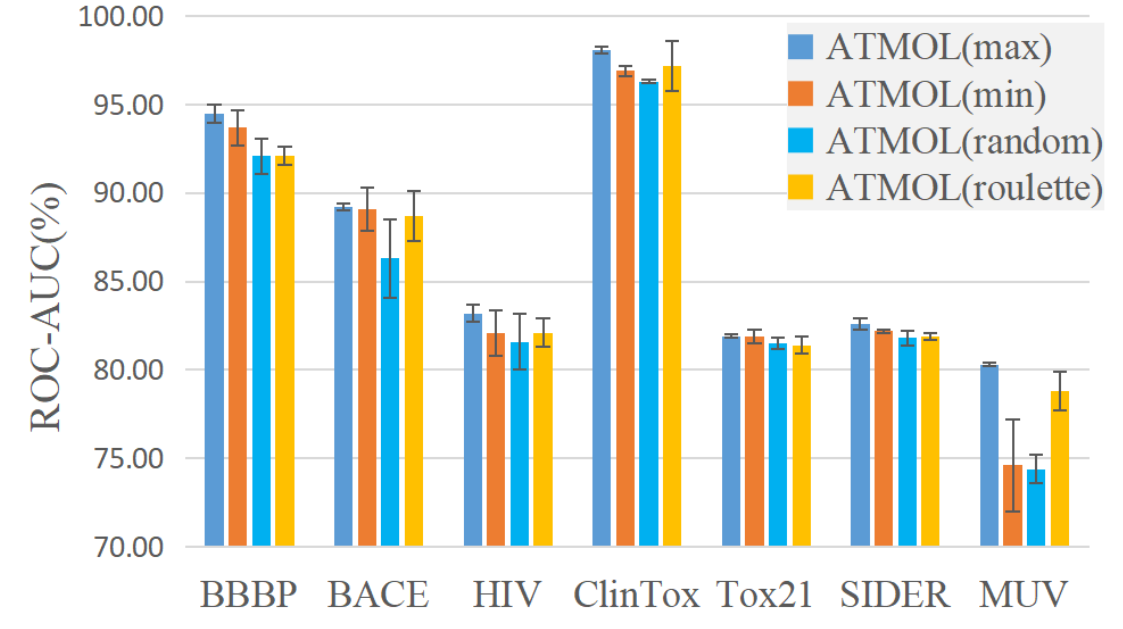}
	\caption{Performance comparison of four masking strategies for graph augmentation on seven molecular property prediction tasks}
	\label{fig:mask_strategy}
\end{figure}

To explore the influence of different graph augmentations on the feature extraction, we compared the performance derived from four masking strategies on seven downstream tasks. As shown in Figure \ref{fig:mask_strategy}, although the prediction performance varied on different molecular properties, the max-weight masking universally achieved the best performance compared to other masking strategies. The random masking had the least performance. Seen by the GAT encoder, masking the nodes and edges with high attention scores produced an augmented graph that differed largely from the positive counterpart sample. So, we drew a conclusion that max-weight masking strategy posed a challenge for the contrastive learning to discriminate a pair of positive samples from a pool of negative samples came from other molecules.

\subsection{Influence of masking rate }
\begin{figure}[htbp]  
	\captionsetup{labelformat=simple, position=top}
	\centering
	\includegraphics[scale=0.5]{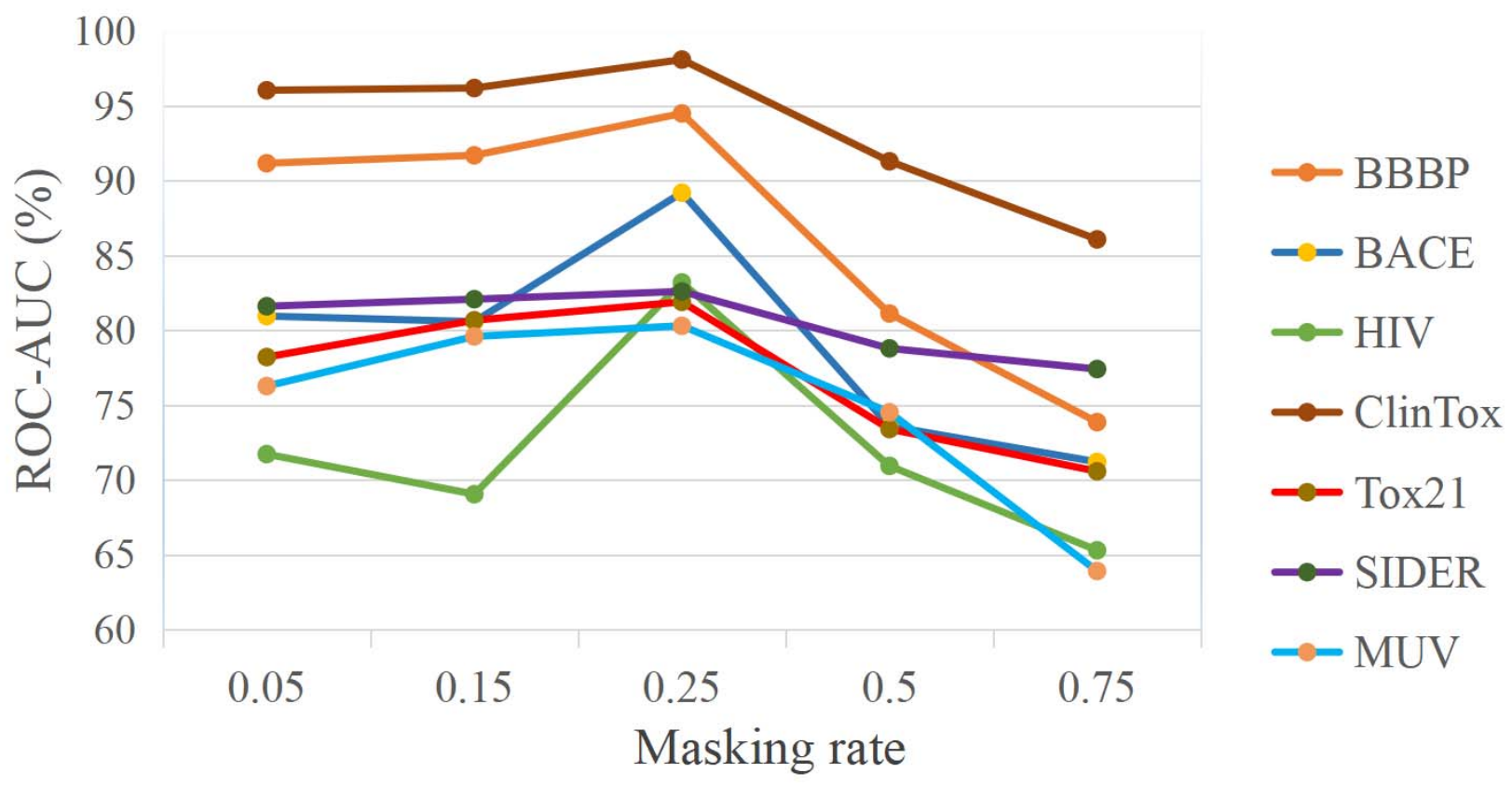}
	\caption{Performance achieved by different masking rate on seven downstream tasks}
	\label{fig:mask_rate}
\end{figure}

We went further to evaluate the impact of masking rate, namely, the percentage of masked nodes (edges). As max-weight masking has been shown to gain best performance, we evaluated its performance when different percentage edges were masked. As shown in Figure \ref{fig:mask_rate}, the masking rate increased from 5\% to 75\%, the performance on seven downstream tasks rose up firstly, and reached the highest ROC-AUC values when masking rate was 25\%. Thereafter, the performance decreased rapidly. Across all seven tasks, we observed similar performance tendency. This result implied that low masking rate did not produce effective augmented molecular graphs, while too high masking rate broke the essential chemical structure so that molecular representation learning was collapsed.

\subsection{Large-scale dataset improve representation learning}
We were interested in whether larger scale of unlabeled data would improve the representation learning or not. So, apart from the in vitro set, we selected 3,000,000 molecules from the ZINC now set as another dataset, which was roughly tenfold of in vitro set. For comparison, we referred them to as small We conducted representation learning on these two datasets separately, and then compared the performance on downstream tasks. As shown in Table \ref{tab:3}, on the large dataset our model achieved higher performance over all downstream tasks. We concluded that self-supervised learning on large-scale dataset yielded molecular representations with better generalizablity.
\begin{table*}[]
\centering
\caption{Performance comparison of small \textit{vs.} large pretraining datasets on seven molecular property by}
\label{tab:3}
\begin{tabular}{ccccccccc} \hline
Dataset & Masking strategy & BBBP & BACE & HIV & ClinTox & Tox21 & SIDER & MUV \\ \hline
\multirow{4}{*}{\textit{\textbf{\begin{tabular}[c]{@{}c@{}}in vitro\\ (small)\end{tabular}}}} &  Max-weight & \textbf{92.1$\pm$0.5} & \textbf{87.3$\pm$0.3} & \textbf{81.2$\pm$0.5} & \textbf{97.5$\pm$0.3} & \textbf{79.1$\pm$0.2} & \textbf{81.9$\pm$0.3} & \textbf{79.1$\pm$0.1} \\
 & Min-weight & 89.4$\pm$0.8 & 82.3$\pm$1.2 & 72.1$\pm$1.3 & 96.9$\pm$0.5 & 77.1$\pm$0.4 & {81.2$\pm$0.1} & 77.6$\pm$1.6 \\
 & random & 83.5$\pm$0.1 & 78.1$\pm$0.2 & 68.9$\pm$0.6 & 96.6$\pm$0.3 & 76.1$\pm$0.2 & 80.8$\pm$0.4 & 76.6$\pm$0.2 \\
 & roulette & 81.5$\pm$0.5 & 78.5$\pm$0.2 & 69.6$\pm$0.2 & 96.2 $\pm$0.1 &75.5$\pm$0.2 & 80.5$\pm$0.2 & 76.5$\pm$0.1 \\ \hline
\multirow{4}{*}{\textit{\textbf{\begin{tabular}[c]{@{}c@{}}now subset\\ (large)\end{tabular}}}} & Max-weight & \textbf{94.5$\pm$0.5} & \textbf{89.2$\pm$0.2} & \textbf{83.2$\pm$0.5} & \textbf{98.1$\pm$0.5} & \textbf{82.5+0.4} & \textbf{82.6$\pm$0.3} & \textbf{80.3$\pm$0.1} \\
 & Min-weight & 93.7$\pm$1.0 & 89.1$\pm$1.2 & 82.1$\pm$1.3 & 96.9$\pm$0.5 & 81.9$\pm$0.4 & 82.2$\pm$0.1 & 74.6$\pm$2.6 \\
 & random & 92.1$\pm$1.0 & 86.3$\pm$2.2 & 81.6$\pm$1.6 & 96.3$\pm$0.3 & 81.5$\pm$0.3 & 81.8$\pm$0.4 & 74.4$\pm$0.8 \\
 & roulette & 91.9$\pm$0.5 & 88.7$\pm$1.4 & 82.1$\pm$0.8 & 97.2 $\pm$0.1 & 81.4$\pm$0.5 & 81.9$\pm$0.2 & 78.8$\pm$1.1 \\ \hline
\multicolumn{1}{l}{} & \multicolumn{1}{l}{} & \multicolumn{1}{l}{} & \multicolumn{1}{l}{} & \multicolumn{1}{l}{} & \multicolumn{1}{l}{} & \multicolumn{1}{l}{} & \multicolumn{1}{l}{} & \multicolumn{1}{l}{}
\end{tabular}

\end{table*}
\subsection{Performance comparison}
To verify the superior performance of our method, we compared it to five other competitive methods. All these methods used self-supervised learning for molecular feature extraction. We concisely introduced the methods as below:
\begin{itemize}
  \item HU. et.al \citep{RN34} pre-trained an expressive GNN at the level of individual nodes as well as entire graphs so as to learn useful local and global representations.
  \item N-Gram \citep{RN35} run node embedding and then constructed a compact representation for the graph by assembling the node embeddings in short walks in the graph.
  \item GROVER \citep{RN36} integrated GNN into Transformer with the context prediction task and the functional motif prediction task.
  \item MolCLR \citep{RN28} proposed a graph contrast learning using graph neural network (GNNs), which generated contrastive pairs by randomly removal of nodes, edges or subgraphs.
  \item MGSSL \citep{RN37} proposed topic-based graph self-supervised learning and a new GNN self-supervised topic generation framework.
\end{itemize}

Table \ref{tab:4} showed the ROC-AUC values of our method and five competitive methods on seven tasks. It was found that our method petrained on the in vitro dataset already outperformed five other methods on all tasks except MUV. Especially, on the SIDER prediction task, our method boosted the performance by 14\% compared to the second best method MolCLR. Moreover, when pretrained on the large-scale dataset, our method achieved greater performance superiority. For example, on ClinTox and Tox21, our method outperformed all other methods by nearly 5\%.
\begin{table*}[]
\centering
\caption{ROC-AUC (\%)of our method and five competitive methods on seven downstream tasks}
\label{tab:4}
\begin{tabular}{lccccccc} \hline
DataSet & BBBP & BACE & HIV & ClinTox & Tox21 & SIDER & MUV \\ \hline
HU. et.al{[}39{]} & 70.8$\pm$1.5 & 85.9$\pm$0.8 & 80.2$\pm$0.9 & 78.9$\pm$2.4 & 78.7$\pm$0.4 & 65.2$\pm$0.9 & 81.4$\pm$2.0 \\
N-Gram{[}40{]} & 91.2$\pm$3.0 & 87.6$\pm$3.5 & 83.0$\pm$1.3 & 85.5$\pm$3.7 & {76.9$\pm$2.7} & 63.2$\pm$0.5 & {81.6$\pm$1.9} \\
Grover{[}41{]} & 68.0$\pm$1.5 & 79.5$\pm$1.1 & 77.8$\pm$1.4 & 76.9$\pm$1.9 & 76.3$\pm$0.6 & 60.7$\pm$0.5 & 75.8$\pm$1.7 \\
MolCLR{[}42{]} & {73.6$\pm$0.5} & {89.0$\pm$0.3} & {80.6$\pm$1.1} & {93.2$\pm$1.7} & 79.8$\pm$0.7 & {68.0$\pm$1.1} & \textbf{88.6$\pm$2.2} \\
MGSSL{[}43{]} & 70.5$\pm$1.1 & 79.7$\pm$0.8 & 79.5$\pm$1.1 & 80.7$\pm$2.1 & 76.5$\pm$0.3 & 61.8$\pm$0.8 & 78.7$\pm$1.5 \\ \hline
ATMOL(small) & 92.1$\pm$0.5 & 87.3$\pm$0.3 & 81.2$\pm$0.5 & 97.5$\pm$0.3 & 79.1$\pm$0.2 & 81.9$\pm$0.3 & 79.1$\pm$0.1 \\
ATMOL(large) & \textbf{94.5$\pm$0.5} & \textbf{89.2$\pm$0.2} & \textbf{83.2$\pm$0.5} & \textbf{98.1$\pm$0.5} & \textbf{82.5+0.4} & \textbf{82.6$\pm$0.3} & 80.3$\pm$0.1 \\ \hline
 & \multicolumn{1}{l}{} & \multicolumn{1}{l}{} & \multicolumn{1}{l}{} & \multicolumn{1}{l}{} & \multicolumn{1}{l}{} & \multicolumn{1}{l}{} & \multicolumn{1}{l}{} \\
 & \multicolumn{1}{l}{} & \multicolumn{1}{l}{} & \multicolumn{1}{l}{} & \multicolumn{1}{l}{} & \multicolumn{1}{l}{} & \multicolumn{1}{l}{} & \multicolumn{1}{l}{}
\end{tabular}

\end{table*}
\subsection{Exploration of model interpretability}
\subsection{Spatial localization of molecular representations}
\begin{figure}[htbp]  

	\captionsetup{labelformat=simple, position=top}
	\centering
	\subfloat[Before training ]
	{
		\begin{minipage}[b]{.5\columnwidth}
			\centering
			\includegraphics[width=4cm]{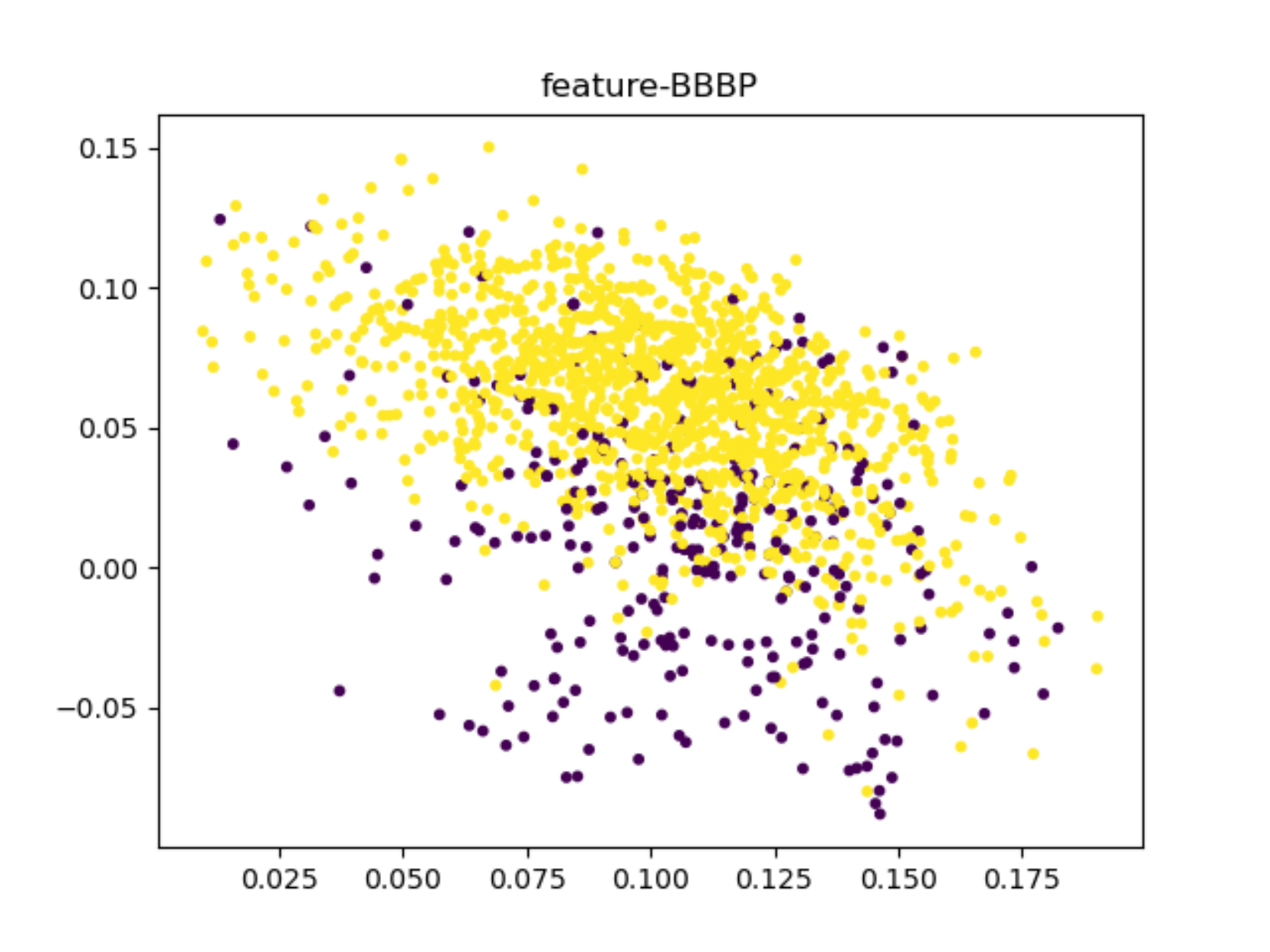}
	\end{minipage}}
	\subfloat[After training]
	{
		\begin{minipage}[b]{.5\columnwidth}
			\centering
			\includegraphics[width=4cm]{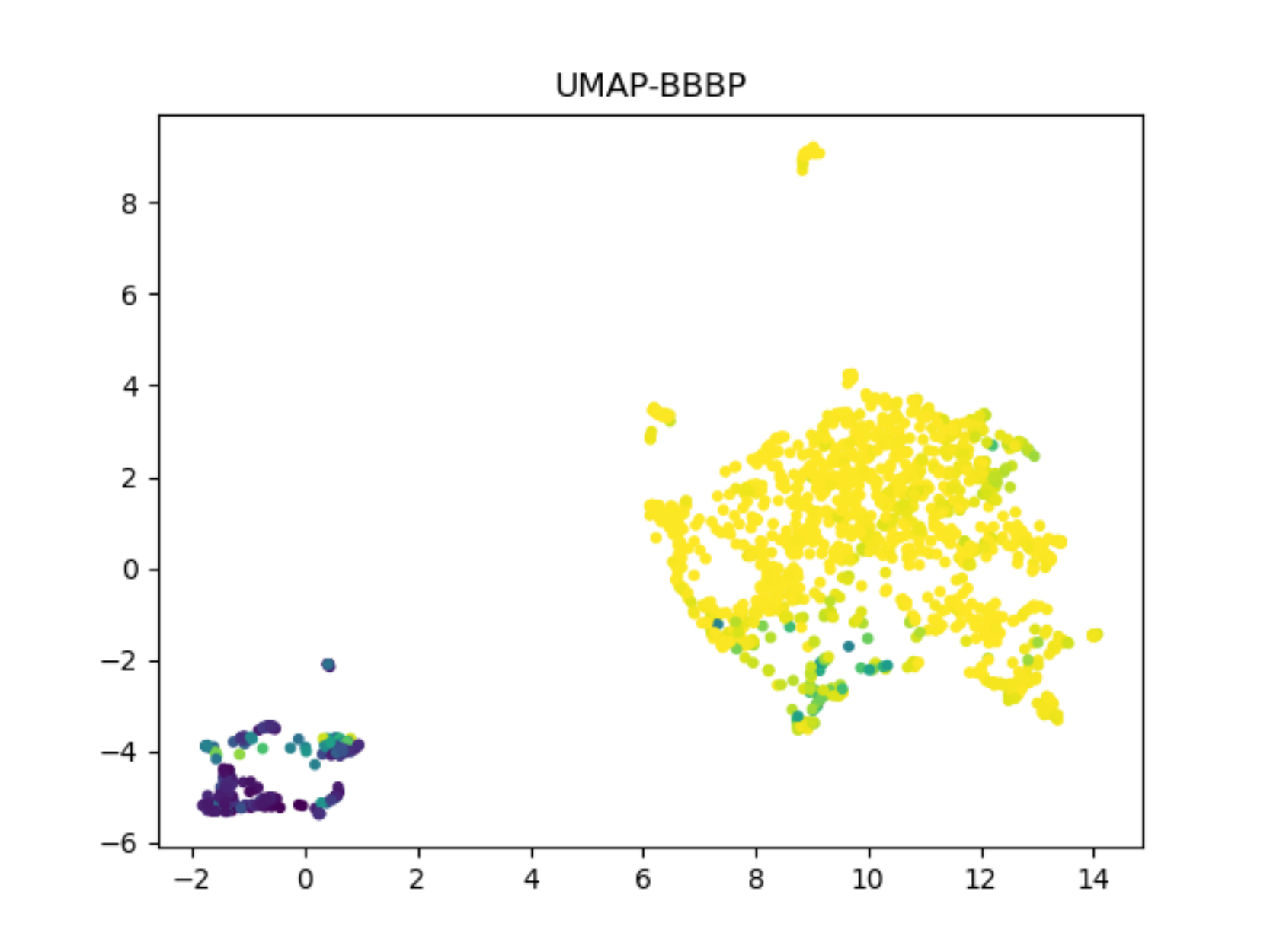}
	\end{minipage}} \\
	\subfloat[Before training]
	{
		\begin{minipage}[b]{.5\columnwidth}
			\centering
			\includegraphics[width=4cm]{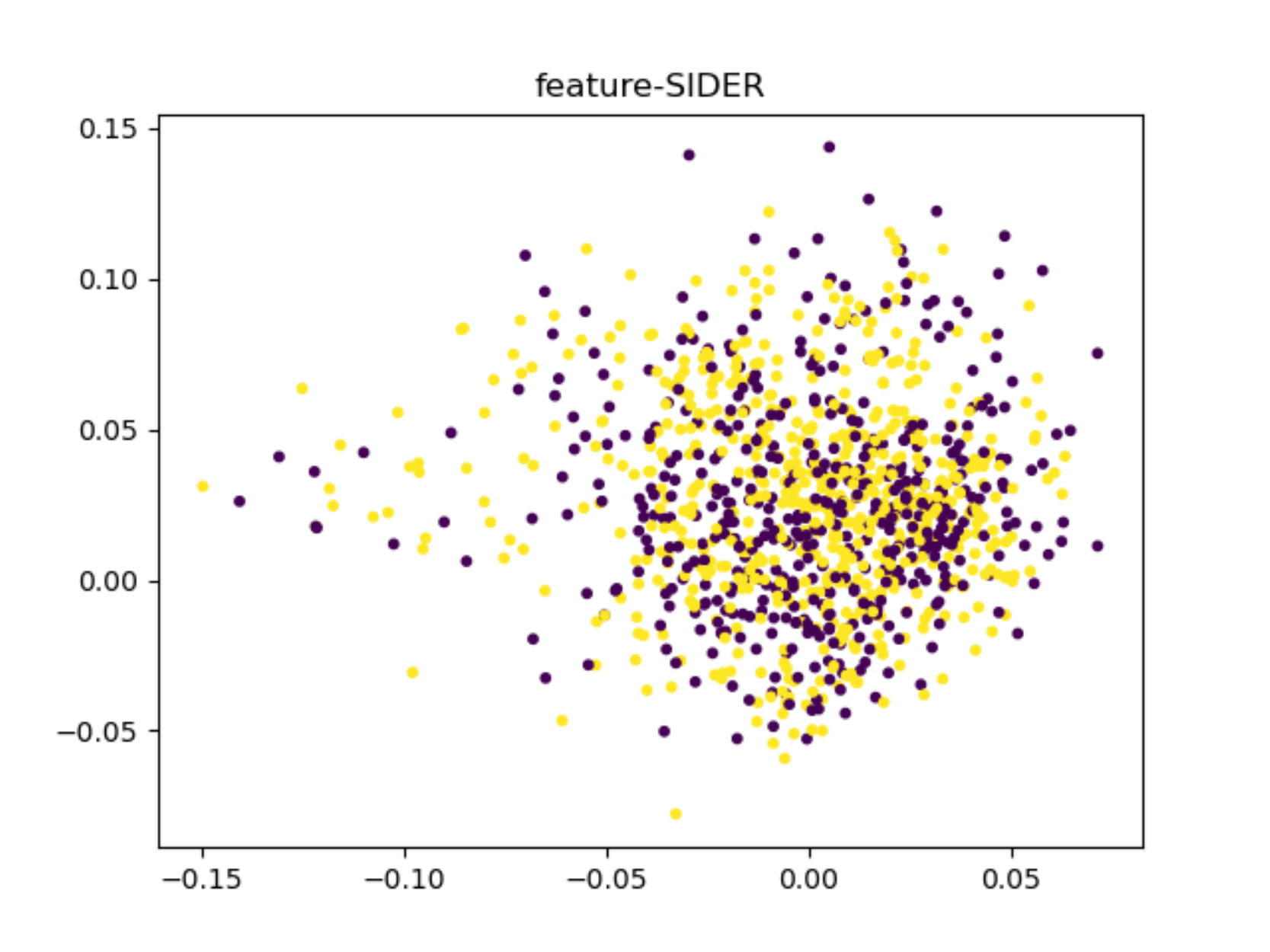}
	\end{minipage}}
	\subfloat[After training ]
	{
		\begin{minipage}[b]{.5\columnwidth}
			\centering
			\includegraphics[width=4cm]{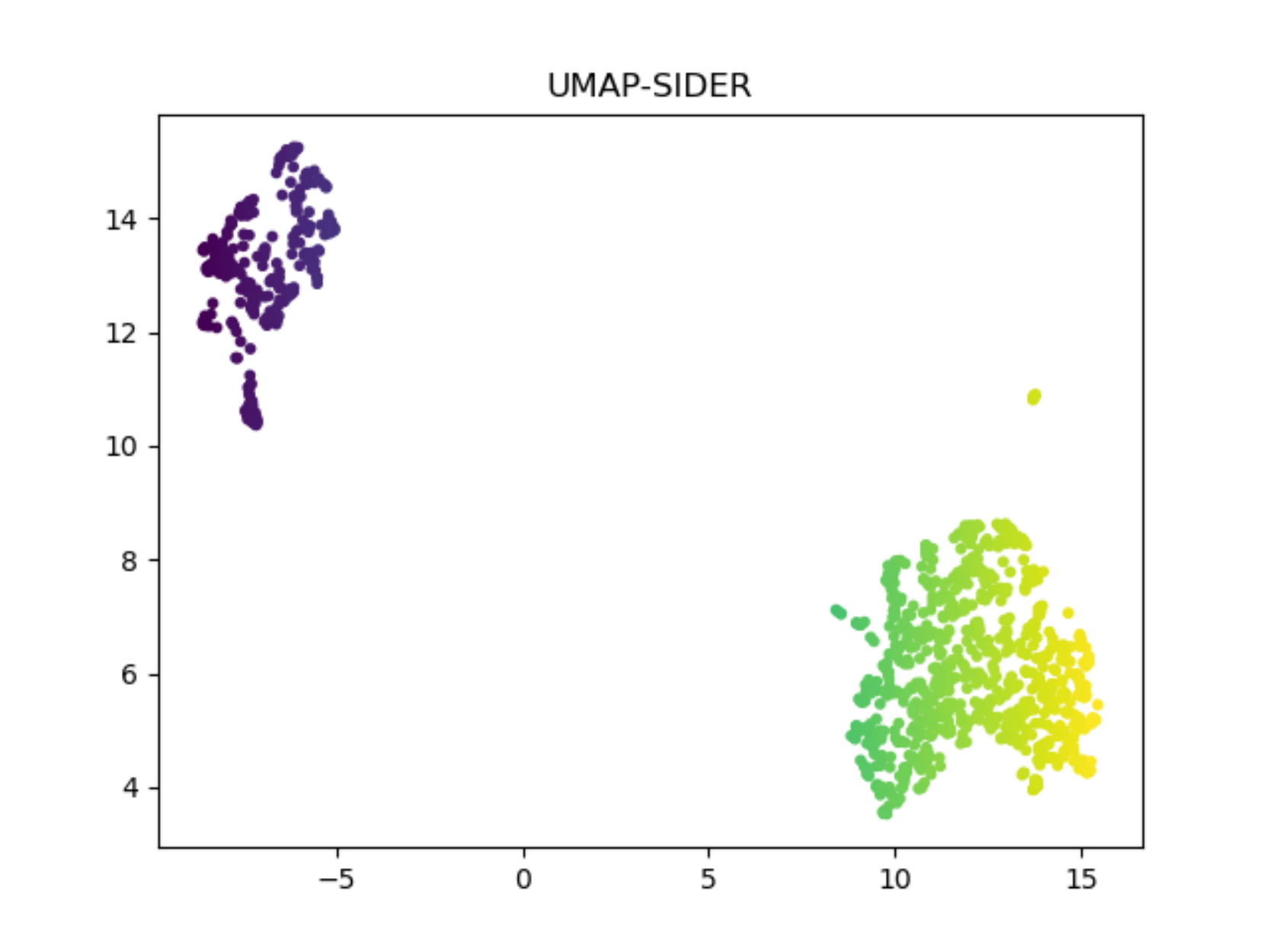}
	\end{minipage}}
	\caption{Spatial localization of tumor area and cells labeled by marker genes. The tumor area was labeled by an experienced pathologist. The localization of T lymphocytes and B lymphocytes was predicted by their marker genes.}
	\label{fig:umap}
\end{figure}

Spatial localization of molecular representations was helpful to verify the effectiveness of contrastive learning. We visualized the molecular representations before and after pretraining by using UMAP tool \citep{RN38}, which is a manifold learning algorithm for dimension reduction with good preservation of data global structure. Figure \ref{fig:umap} showed the 2D embeddings of molecules in BBBP and SIDER sets. The initial molecular representations spatially distributed in confusion, while after pretraining these molecules belonging to one same class gathered together and separated from other classes. The observation illustrated the our method can effectively detect the physicochemical properties from chemical structure, so that the molecules with similar physicochemical properties gained similar latent representations.

\subsection{Attention weight revealed important chemical substructure}
\begin{figure*}[htb]
	\centering
	\includegraphics[scale=0.55]{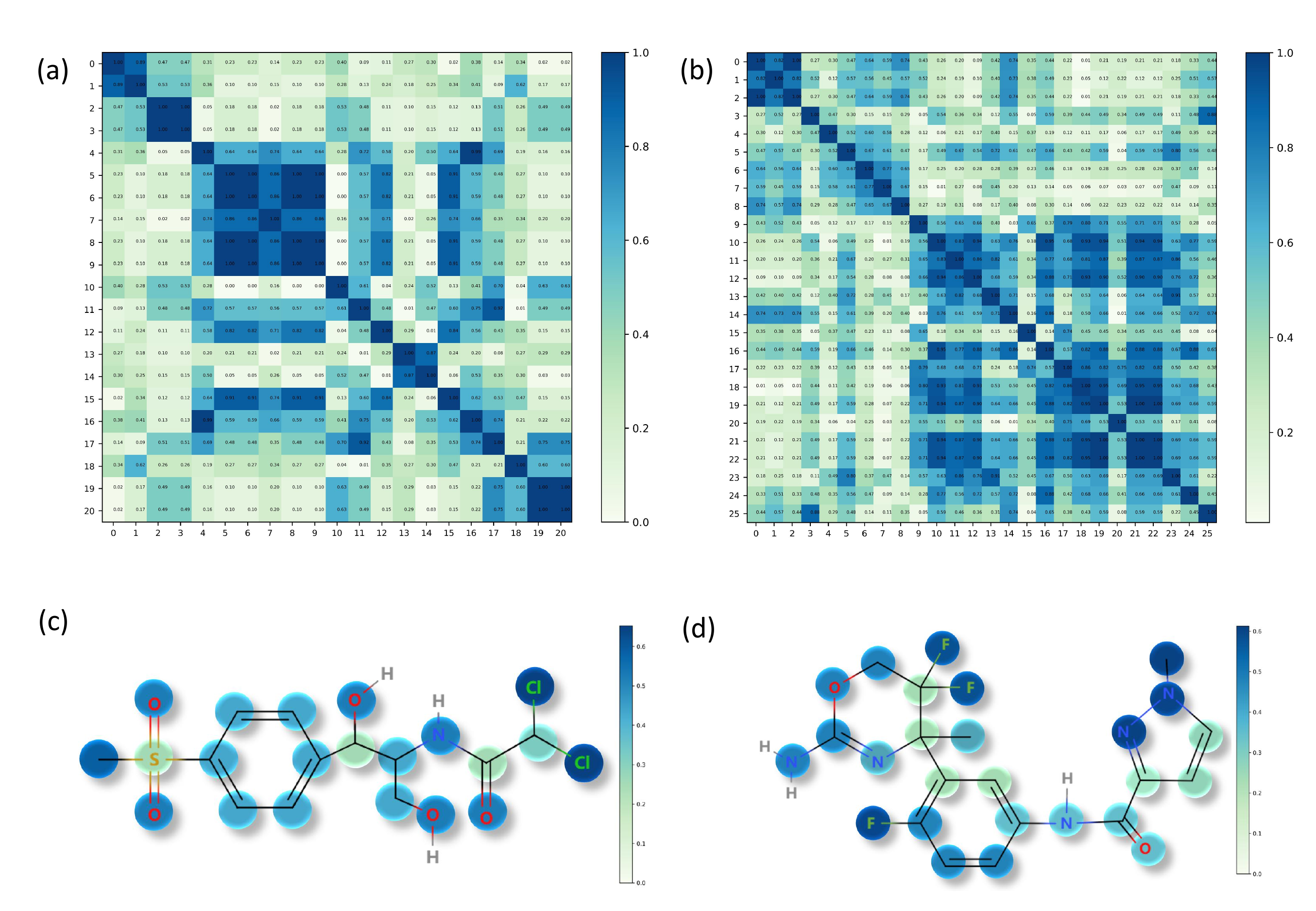}
	\caption{Visualization of correlation heatmaps of attention weights and molecule structure with atoms colored by attention weights. (a) and (c) showed the exemplar molecule selected from BBBP set, (b) and (d) showed the exemplar molecule selected from BACE set.}
	\label{fig:heatmap}
\end{figure*}

In the molecular representation learning phase, we constructed molecular graphs from the SMILES descriptions, and then used graph attention network to identify chemical components that were important to specific prediction tasks. To investigate how the attention mechanism affected the focus of representation learning, we unfreeze the attention parameters and fine-tuned them during transfer learning. For intuitive interpretation, we visualized the attention weights in the molecular graph. From the BBBP dataset \citep{RN39} regarding to membrane permeability, we randomly selected a molecule as an exemplar, whose SMILES is C[S](=O)(=O)c1ccc(cc1)[C@@H](O)[C@@H](CO) NC(=O)C(Cl)Cl. We computed the Pearson correlation coefficients of attention weights for each pair of atoms, visualized the correlation matrix as a heatmap. As shown in Figure \ref{fig:heatmap}a, The heatmap revealed several highly correlated atomic groups, indicating that they functioned together to affect specific molecular properties. We further observed that the benzene ring in this molecule may played important roles in determining the membrane permeability. Similarly, we randomly select a molecule FC1(F)COC(=NC1(C)c1cc(NC(=O)c2nn(cc2)C)ccc1F)N from the BACE dataset \citep{RN40}, its heatmap also illustrated several atomic groups, as shown in Figure \ref{fig:heatmap}b.

To further explore the influence of individual atoms on specific molecular properties, we visualized atomic-level attention weights. As the BBBP task focused on the membrane permeability, we found that two Cl atoms in the exemplar molecule had high attention weight, as shown in Figure \ref{fig:heatmap}c. Because the Cl atom has a strong electron attraction, we assume it affects the polarity of the molecule to a large extent, thereby affecting the membrane permeability. Also, since the hydroxyl groups promotes hydrophilicity, and we accordingly found that the hydroxyl groups were given relatively high attention \citep{RN41}.

Similarly, another exemplar molecule selected from BACE dataset is a human beta-secretase 1 (BACE-1) inhibitor. According to precious study by Luca G. Mureddu et al \citep{RN42}. the heterocytosine aromatic family had the inhibitory effect to BACE-1. As shown in Figure \ref{fig:heatmap}d, the isocytosine component of this molecule has received more attention. The visualization and interpretability exploration illustrate how our model paid attention to relevant features from the perspective of molecular property prediction tasks.
\section{Discussion and Conclusion}
The diversity of negative samples has been shown to greatly affect representation learning. There are two main methods to construct negative samples for contrastive learning. Some methods maintained a negative sample queue \citep{RN23} and iteratively updated it in a FIFO manner, while other methods used only the samples in the current mini-batch as negative samples \citep{RN24}. In our study, beyond the samples in minibatch as negative samples, we added the augmented molecule graphs generated by attention-wise masking strategy to the negative sample queue, so that the negative samples was greatly extended and diversified.

Our attention-wise mask of molecular graph generated different contrastive views. By comparison, we found that the contrastive views generated by simultaneously masking edges and nodes achieved best performance in almost all downstream tasks, except for Tox21 and MUV. This was consistent with the conclusions of MolCLR[28].

Moreover, we found that max-weight masking, i.e. masking the edges or nodes with large attention weights, achieved best performance. Intuitively, max-weight masking is similar to the idea of adversarial learning, by which each time an augmented molecular graph was generated with the largest difference to the positive couterpart view, from the perspective of contrastive loss. Moreover, the graph augmentation process dynamically tracked the change of attention weights, so that our model was forced to inspect different components of the molecular graph and finally reach steady state. Therefore, we concluded that the challenging contrastive views were helpful to learn important semantic structure.

In summary, our self-supervised representation learning on large-scale unlabeled molecules significantly improved the performance of various molecular property prediction tasks. This a task-agnostic pretraining and thus yielded the molecular representation with desirable expressiveness and generalizability.

\section{Author contributions statement}
L.H. and Y.H. conceived the main idea and the framework of the manuscript. L.H. and Y.H. performed the experiments. X.L. and L.D. helped to improve the idea. H.L. wrote the manuscript. H.L. supervised the study and provided funding. All authors read and commented on the manuscript.

\section{Data availability}
Source code and all datasets used in this study are available at \\ \href{https://github.com/moen-hyb/ATMOL}{https://github.com/moen-hyb/ATMOL}

\section{Acknowledgments}
This work was supported by the National Natural Science Foundation of China under grants No.~62072058 and No.~61972422.

\bibliographystyle{unsrt}
\bibliography{reference}



\begin{biography}{}{\author{Hui Liu} is a professor at School of Computer Science and Technology, Nanjing Tech University, Nanjing, China. His research interests include Bioinformatics and Deep Learning.}
\end{biography}
\begin{biography}{}{\author{Yibiao Huang} is an master student at School of Computer Science and Engineering, Central South University, Changsha, China.}
\end{biography}

\begin{biography}{}{\author{Xuejun Liu} is a professor at School of Computer Science and Technology, Nanjing Tech University, Nanjing, China. His research interests include data mining and deep learning.}
\end{biography}
\begin{biography}{}{\author{Lei Deng} is a professor at School of Computer Science and Engineering, Central South University, Changsha, China. His research interests include data mining, bioinformatics and systems biology.}
\end{biography}

\end{document}